\documentclass[preprint,tightenlines,showpacs,preprintnumbers,amsmath,amssymb,prb]{revtex4}

\usepackage{graphicx}
\usepackage{bm}

\begin{document}
\draft
\title{Single-walled carbon nanotube bundle under hydrostatic pressure studied by the first-principles calculations}

\author{Xiaoping Yang}
\email{bunnyxp@hotmail.com}
\affiliation{Group of Computational
Condensed Matter Physics, National Laboratory of Solid State
Microstructures and Department of Physics, Nanjing University,
Nanjing 210093, P. R. China} \affiliation{Department of Physics,
Huainan Normal University, Huainan, Anhui 232001, P. R. China}

\author{Gang Wu}
\affiliation{Group of Computational Condensed Matter Physics,
National Laboratory of Solid State Microstructures and Department
of Physics, Nanjing University, Nanjing 210093, P. R. China}

\author{Jian Zhou}
\affiliation{Group of Computational Condensed Matter Physics,
National Laboratory of Solid State Microstructures and Department
of Physics, Nanjing University, Nanjing 210093, P. R. China}

\author{Jinming Dong}
\email[Corresponding author. Email address: ]{jdong@nju.edu.cn}
\affiliation{Group of Computational Condensed Matter Physics,
National Laboratory of Solid State Microstructures and Department
of Physics, Nanjing University, Nanjing 210093, P. R. China}

\begin{abstract}
The structural, electronic, optical and vibrational properties of
the collapsed (10,10) single-walled carbon nanotube bundle under
hydrostatic pressure have been studied by the first-principles
calculations. Some features are observed in the present study:
First, a collapsed structure is found, which is distinct from both
of the herringbone and parallel structures obtained previously.
Secondly, a pseudo-gap induced by the collapse appears along the
symmetry axis \textit{$\Gamma $X}. Thirdly, the relative
orientation between the collapsed tubes has an important effect on
their electronic, optical and vibrational properties, which
provides an efficient experimental method to distinguish
unambiguously three different collapsed structures.

\end{abstract}


\maketitle

\begin{large}
\begin{center}
{\bf I. Introduction}
\end{center}
\end{large}

In the past decade carbon nanotubes (CNTs) [1-3], both
single-walled (SWNT) and multi-walled (MWNT), had been extensively
investigated due to their special electrical and mechanical
properties, as well as their potential applications in future
nanostructured materials, such as nanoscale quantum wires, single
electron and field-effect transistors and sensors.

It is well known that the physical properties of the CNTs depend
much on their geometrical structures, and so can be easily changed
by an applied pressure or strain, which could be used to fabricate
the nanoscale electromechanical coupling devices and transducers.
For example, a uniaxial strain on the SWNTs can cause a
metal-semiconductor transition. On the other hand, both of the
SWNT and MWNT bundles under hydrostatic pressure [4-15] have been
studied experimentally and theoretically, showing a structural
phase transition (SPT) at a critical pressure. It is found that
the Raman peaks shift to higher frequencies with increasing
hydrostatic pressure, and the radial breathing mode (RBM)
disappears from the spectrum above the critical pressure [5,7].
Very recently, Elliott \textit{et al.} [13] reported their
classical molecular dynamics simulations on the SWNT bundles under
hydrostatic pressure, which can collapse to a herringbone
structure after the SPT, and their Raman spectrum measurement
showed an excellent agreement with their simulations. However,
Zhang \textit{et al.} [16] put forward that the stress-induced
herringbone structure is not the most favorable in energy, but
found a parallel array of the collapsed nanotubes which is even
more favorable. Therefore, it is unclear up to now which one, the
parallel or the herringbone, is the most favorable structure of
the SWNT bundles above the critical hydrostatic pressure.

It is well known that the optical spectroscopy, from the Raman and
infrared modes to the visible optical absorption spectrum, is a
powerful experimental method to detect the geometrical structures
of the systems, because they are sensitive to the changes of the
geometrical structures and electronic structure of the materials,
especially in the nanostructured materials, such as CNTs. So, in
this paper, we systematically investigate electronic, optical and
vibrational properties of the different collapsed structures for
the (10,10) SWNT bundles under hydrostatic pressure using the
first-principles method, in order to identify them unambiguously.

The paper is organized as follows: In Sec. II, we introduce the
employed method. Then, the calculated results and discussions are
given in Sec. III. Finally, some concluding remarks are offered in
Sec. IV.

\begin{large}
\begin{center}
{\bf II. Method}
\end{center}
\end{large}

We firstly study the changes of geometrical structure driven by a
combination of hydrostatic pressure and van der Waals forces. The
zero-temperature structural minimizations of the enthalpy
($H=U+PV$) were carried out on a supercell containing a 2x2x2
(10,10) SWNT bundle using the universal force field (UFF) method
[17,18]. We only change the diagonal part of pressure tensor,
which ensure the hydrostatic pressure is applied. In order to
induce a structural phase transition, a step-wise monotonically
increasing hydrostatic pressure was applied to the (10,10) SWNT
bundle at zero temperature. The pressure was increased from 0 to 5
GPa in 50 steps, minimizing the enthalpy of the SWNT bundle after
each pressure increment. The geometrical structures of the
collapsed SWNT bundle were optimized again employing conjugate
gradient technique through the first-principles method [19], and
in the final geometry no forces on the atoms exceed 0.001 eV/\AA.
However, no qualitative structure difference is found after the
further first-principles geometry optimization. Then the total
energy plane-wave potential method [19] in the framework of local
density approximation (LDA) has been used to investigate the
structural, electronic, optical and vibrational properties of the
collapsed SWNT bundle under hydrostatic pressure. The ion-electron
interaction was modeled by the projector augmented wave (PAW)
potentials with a uniform energy cutoff of 500 eV.  The maximum
spacing between K points was 0.03 \AA $^{-1}$ and the smearing
width was taken to be 0.04 eV in the ground state. The Raman
intensity was calculated by combining the first-principles
calculations and the empirical bond polarizability model
(EBPM)[20,21].

\begin{large}
\begin{center}
{\bf III. Results and Discussions}
\end{center}
\end{large}

Taking the example of collapse to the parallel structure, the
volume change was measured as a function of the applied
hydrostatic pressure using the UFF method, given in Fig. 1(a),
from which a discontinuous SPT can be clearly seen at about 3 GPa.
In Fig. 1(b)-(d) shown are three different collapsed structures of
the (10,10) SWNT bundle after the SPT at 3 GPa, among which the
parallel structure [Fig. 1(b)] and the herringbone structure [Fig.
1(d)] are in good accordance with the classic molecular dynamics
simulation results, obtained previously in Ref.[16] and Ref.[13],
respectively. It can be seen that the SWNTs spontaneously collapse
to those with a dumbbell-like cross section, where the separation
between two opposite parallel walls is approximately equal to the
distance between layers in turbostratic graphite. However, we have
also found an in-between configuration [Fig. 1(c)], which is an
almost degenerate in enthalpy with the parallel one, having a
difference of only 0.3 meV per atom (see the following Table I).
It should be noted that all tubes are equivalent to each other in
a unit cell of the parallel structure, but are different in other
two structures, in which there are two inequivalent tubes in their
one unit cell. For comparison, we list in Table I all our LDA
results of the three collapsed structures, from which we can find
both of the parallel and in-between structures are more favorable
in enthalpy than the herringbone one, but the largest difference
of the enthalpy per atom in the three structures is only about 2.4
meV. So, it is very important to identify them clearly in future
experiments.

The electronic band-structure along the nanotube symmetry axis
\textit{$\Gamma $X}, for an individual (10,10) SWNT and its bundle
at 0 GPa, and the three different collapsed structures at 3 GPa
are calculated by the LDA method and given in Figs. 2(a)-2(e),
respectively. Here \textit{k}-points sampling of 1x1x120 is used
in our calculations. The intertube van der Waals forces and the
structure deformation break the rotational symmetry of the SWNT
even at zero pressure, making its energy bands near Fermi Level
split, as shown in Fig. 2(b), which is different from that of an
individual SWNT [Fig. 2(a)] at the same zero pressure. Also, the
density of states (DOS) for the three collapsed structures has
also been calculated with a \textit{k}-points sampling of 0.02 \AA
$^{-1}$ in the Brillouin Zone (BZ), and shown in Fig. 3. From a
comparison between Fig. 2 and 3, it is found that although the
three collapsed structures always remain in metal [as seen in Fig.
3], there exists still a basic difference between their band
structures. For example, the band structure along the
\textit{$\Gamma $X} direction [Fig. 2(c)] of the parallel
structure shows a pseudo-gap of 0.2 eV at the Fermi Level, and
another small one of 0.08 eV emerges in the electronic band of the
herringbone structure [Fig. 2(e)]. However, no the pseudo-gap
exists in the band of the in-between structure, in which the top
of the valence band and the bottom of the conduction band lie
separately at different \textit{k} points without the pseudo-gap
in Fig. 2(d). In fact, this kind of band differences between the
three collapsed structures comes from the different weak
dispersions along the directions perpendicular to the tube axis,
caused by the different numbers and relative orientation of the
inequivalent tubes in their unit cells.

The changes of the electronic structure can be embodied well in
their optical absorptions, which are calculated with the
\textit{k}-points sampling of 1x1x120 for the individual SWNT and
of 0.02 \AA $^{-1}$ in the BZ for other bundle structures, and
shown in Figs. 4(a)-4(e). The individual (10, 10) SWNT [Fig. 4(a)]
has six absorption peaks polarized along tube direction (Z
direction), lying at 1.591, 2.788, 3.516, 3.903, 11.92 and 14.069
eV, respectively, which are almost the same as those in Ref [22].
For the bundle at zero pressure, the original characteristic peaks
of the individual SWNT shown in Fig. 4(a) are slightly shifted
because its band splitting caused by the tube-tube interaction in
the bundle, as seen clearly in Fig. 4(b). In addition, a strong
absorption peak polarized along tube direction emerges at 0.184 eV
due to the existing pseudo-gap along the direction parallel the
\textit{$\Gamma $X}, which had been predicted by Paul Delaney
\textit{et al.} [23] and further confirmed by a following
experiment [24]. To investigate the evolution of this strong
absorption peak and the original six ones with the applied
hydrostatic pressure, we label them with symbols of '0', '1', '2',
'3', '4', '5' and '6', respectively, in Fig. 4, and list their
calculated energy positions in Table II.

It is seen from Fig. 4 and Table II that: 1) the original distinct
peaks of No. '4' and '5' for the undeformed (10,10) SWNT bundle in
Fig. 4(b) become indistinct or disappear in Figs.4(c)-4(e) for
three collapsed structures, and other characteristic peaks in
Figs. 4(b) are blue-shifted or red-shifted. 2) The peak of No. '0'
is blue-shifted from 0.184 eV in Fig. 4(b) to 0.354 eV in Fig,
4(c) for the parallel structure and to 0.53 eV in Fig. 4(e) for
the herringbone one. But, it makes a slightly red-shift to 0.15 eV
in Fig. 4(d) for the in-between structure. 3) The peak of No. '1'
is red-shifted for the parallel and the in-between structures, as
shown in Fig. 4(c) and 4(d), but splits into three distinctive
peaks at 1.149, 1.584 and 1.965 eV for the herringbone structure,
respectively, seen in Fig. 4(e). It seems that all the above
characteristics of the optical absorption, especially their lower
frequency peaks, could be used to identify experimentally three
different collapsed structures.

On the other hand, it can be found from Figs. 4(b)-4(e) that there
exists a large difference in the optical absorption anisotropy for
the undeformed (10,10) SWNT bundle and other three collapsed
configurations, which is listed as follows: 1) First of all, the
optical responses polarized along X and Y directions are almost
the same in magnitude in the undeformed and the herringbone
structures, but much different in the parallel and the in-between
ones, which could be caused by the relative orientations between
the nearest tubes in them. 2) In addition, it can be found that
the optical responses along X direction for both of the parallel
and in-between ones are almost the same in magnitude too, and in
the low-energy region weaker than that polarized along Y
direction, because in both structures the normal of the flat
segment of the collapsed tubes makes an almost the same angle with
X or Y directions. The flat segment of the collapsed dumbbell-like
cross section [see Fig. 1(b)] looks very similar to the graphite
sheet. It is well known that in graphite, the optical absorption
component polarized perpendicular to the carbon-layer is much
smaller than that parallel to the layer [see Ref. 22]. It is seen
clearly from Fig. 1(b) and 1(c) that the Y component of the
optical absorption in both parallel and in-between structures has
a more parallel component of the graphite than the X component,
which is finally reflected in Fig. 4(c) and 4(d). We have repeated
the same calculations for other SWNT bundles with the larger
radius [for example, the armchair (12,12), (14,14), (16,16),
(18,18) and (20,20) tubes], and found the flat segments in their
cross sections become the larger and larger with increasing tube
radius, indicating the graphite-like characteristic anisotropy
would be more notable in the collapsed SWNT bundles with larger
radii. On the other hand, this character is absent in the optical
spectra of the herringbone configuration because as seen from Fig.
1(d), its X and Y directions form an almost the same angle with
the flat segment of the collapsed tubes, making so the X and Y
components of its optical absorptions contain almost the same
amount of the parallel component of the graphite.

Now, let us discuss their vibrational properties. It is known that
the RBM is the most important low-frequency mode in the Raman
spectra of a perfect SWNT, in which all carbon atoms are subject
to an in-phase radial displacement. It was found that the RBM
frequency of a SWNT is independent of its chiral angle and depends
only on its diameter in an inverse proportion. Here, we have
calculated by the EBPM the nonresonant Raman spectra of a (10 10)
SWNT bundle at 0 GPa and its three different collapsed structures
at 3 GPa, and shown them in Fig. 5. Because there is no obvious
difference between the Raman spectra of three collapsed structures
in high-energy region, only the Raman peaks in the range from 0 to
350 cm$^{ - 1}$ are given in Fig. 5. From Fig. 5(a) and its inset,
we can find the RBM of the individual SWNT moves from 174 cm$^{ -
1}$ to the higher energy 205 cm$^{- 1}$ in the undeformed bundle
due to the van der Waals forces. Meantime, its low-frequency peak
at about 100 cm$^{ - 1}$ seen in the inset of Fig. 5 (a) splits,
producing several peaks in Fig. 5(b). After the SPT, it is clearly
seen from Figs. 5(b)-5(d) that the Raman spectra of three
collapsed structures have been changed greatly: (1) the whole
Raman spectra extend more widely than before the SPT; (2) more
Raman peaks emerge, e.g., the big distinctive peak at about 300
cm$^{ - 1}$; (3) Most importantly, the Raman peaks shift to higher
frequencies. We made a search for the Raman frequency range from
180 to 350 cm$^{- 1}$ by detailed analysis on the vibration modes
of the Raman peaks, but no RBM-like mode is found, i.e., the
original RBM of the undeformed structure disappears from this
frequency region, which is well consistent with the experimental
results [5,7].

All these changes in the Raman spectra are caused by the large
structural difference before and after the SPT, leading to much
different normal vibration modes after the SPT. Specifically, by
detailed analysis on the vibration modes shown in Fig. 5(b) and
5(d), two main differences can be found in them: firstly a
distinct Raman peak at 177 cm$^{-1}$ (quadrupole vibration mode)
in the parallel structure disappears from the corresponding
spectrum range of the herringbone structure, in which, however,
the quadrupole vibrations induce two strong Raman-active peaks at
115 and 136 cm$^{-1}$ due to coupling with other vibration modes
in low-energy region; and secondly some Raman peaks appear below
100 cm$^{-1}$ in the herringbone structure, which should be mainly
produced by the relative vibrations between two inequivalent tubes
in its unit cell. In order to illuminate these vibration modes, we
show the atomic displacements at 177 cm$^{-1}$ for the parallel
structure and 81, 115 and 136 cm$^{-1}$ for herringbone one in
Figs. 6(a) and 6(c), respectively. The Raman peaks below 100
cm$^{-1}$ also exist in that [Fig. 5(c)] of the in-between
structure, but its Raman spectra above 100 cm$^{-1}$ is very
similar to that of the parallel structure with only a slight
frequency shift, e.g. the quadrupole vibration mode at 177
cm$^{-1}$ in Fig. 5(b) shifts to present 167 cm$^{-1}$. For
comparison, we also show its atomic displacements at 64 and 167
cm$^{-1}$ in Fig. 6(b). So, our results indicate that the three
collapsed structures can also be distinguished clearly by their
low frequency Raman-active modes.

Experimentally, SWNT bundles contain tubes with different
diameters or chiralities, forming the so-called "mixed" bundles in
real sample, which seems to make difficult to compare the
experimental data with our calculated results on the pure (10,10)
bundle because unfortunately, at the present time, it is not
possible to make a numerical simulation on the "mixed bundle"
under hydrostatic pressure, using both classical mechanics and the
first-principles method. Even so, however, our calculation results
in this paper can still be helpful for analyzing the experimental
data on the bundle.

In general, the Raman spectrum of the vibrational mode and x-ray
diffraction of the tube bundles are often used to identify tubes'
diameters and their chiral angles in the real sample. The X-ray
analysis could reveal the average diameter of the tubes in the
sample, and the Raman spectrum of the breathing mode of the bundle
can also fix the diameter of the SWNTs in the bundle. Based upon
all these data, we can identify rather accurately what kinds of
SWNTs exist in the bundle by a comparison of the data with those
of theoretical calculations. After that, we need to compare the
experimental data, e.g., the SPT, the optical absorption and Raman
spectrum, of the "mixed" SWNT bundle under pressure with those
obtained from the theoretical calculations, which, of course,
should be done for all possible pure SWNT bundle with different
kinds of SWNTs in the real bundle sample. So, it would have no big
problem for the experimentalist to relate their data directly to
the theoretical analysis on the pure bundle structure. Usually,
the (10,10) tube is mostly grown in the real sample. That is why
in this paper we made only calculation on the (10,10) tube bundle.

\begin{large}
\begin{center}
{\bf IV. Summary}
\end{center}
\end{large}

In summary, we systematically investigate the structural,
electronic, optical and Raman-active properties of the collapsed
(10,10) SWNT bundle under hydrostatic pressure. We have found an
in-between configuration under the collapse pressure, which is
distinct from both the parallel and herringbone structures found
previously, but is almost the degenerate in enthalpy with the
parallel structure. In addition, we find the collapse possibly
induces the pseudo-gap along tube axis \textit{$\Gamma $X}, and
the relative orientation between the collapsed tubes has an
important effect on the electronic structure, optical and
vibrational properties. Furthermore, our results indicate three
different collapsed structures can be distinguished clearly by
their electronic, optical and Raman-active properties.

\begin{acknowledgments}
This work was supported by the Natural Science Foundation of China
under Grant No. 10474035 and A040108, and also by the State Key
program of China through Grant No. 2004CB619004.
\end{acknowledgments}

\newpage

\begin{center}
\textbf{TABLE}
\end{center}

\begin{table}[htbp]
\textbf{Table I.} Total energies (E) and enthalpies (H) of the
parallel, the new in-between, and the herringbone structures at
the collapsed pressure of 3 GPa and zero temperature.
\begin{tabular}
{|p{100pt}|p{100pt}|p{100pt}|p{100pt}|}
\hline Stucture &E/atom (eV)& PV/aotm (eV)& H/aotm (eV)\\
\hline Parallel & -9.9804 & 0.1639 & -9.8165 \\
\hline New in-between & -9.9809  & 0.1647  & -9.8162 \\
\hline Herringbone & -9.9839  & 0.1698  & -9.8141 \\
\hline
\end{tabular}
\label{tab1}
\end{table}

\begin{table}[htbp]
\textbf{Table II.} The calculated energy positions of the
characteristic optical absorption peaks [labelled by symbol of
'0', '1', '2', '3', '4', '5' and '6'] polarized parallel to the
tube axis (Z direction) in Fig. 4(a)-4(e).
\begin{tabular}
{|p{120pt}p{38pt}p{90pt}p{38pt}p{38pt}p{38pt}p{38pt}p{38pt}|}
\hline character peak&0&1&2&3&4&5&6\\
\hline Stucture &&&Energy(eV)&&&&\\
\hline Individual (0 GPa) && 1.591& 2.788&3.516 & 3.903& 11.920& 14.069\\
\hline Bundle (0 GPa)& 0.184& 1.544& 2.836&3.311 & 3.869& 11.941& 14.022 \\
\hline Parallel (3 GPa)&0.354& 1.367&2.434& 3.318&4.155&12.097&14.423\\
\hline New in-between (3 GPa)& 0.150& 1.516&2.768& 3.291&3.992&12.009 &14.443 \\
\hline Herringbone (3 GPa)&0.53 & 1.149,1.584,1.965&2.904& 3.339&3.788&12.118& 14.409 \\
\hline
\end{tabular}
\label{tab1}
\end{table}

\newpage
\begin{center}
\textbf{Figure Captions}
\end{center}

\begin{figure}[htbp]
\includegraphics[width=0.8\columnwidth]{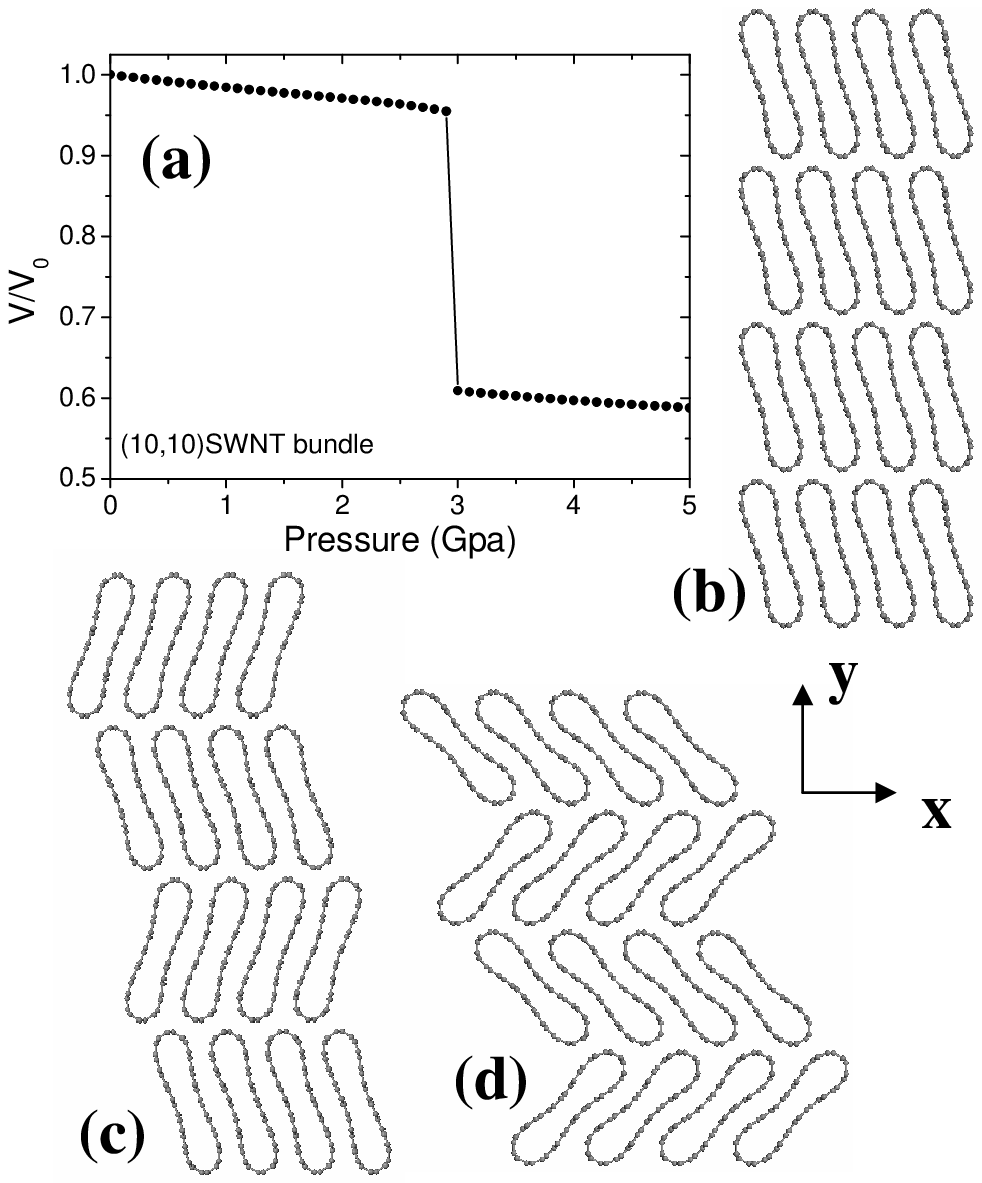}
\label{fig1} \caption{Color online)(a) Loading curve for (10,10)
SWNT bundle as a function of hydrostatic pressure. Snapshots of the
cross section of a (10,10) SWNT bundle at 3 GPa: (b) the parallel
structure, (c) the in-between structure, and (d) the herringbone
structure.}
\end{figure}

\begin{figure}[htbp]
\includegraphics[width=0.8\columnwidth]{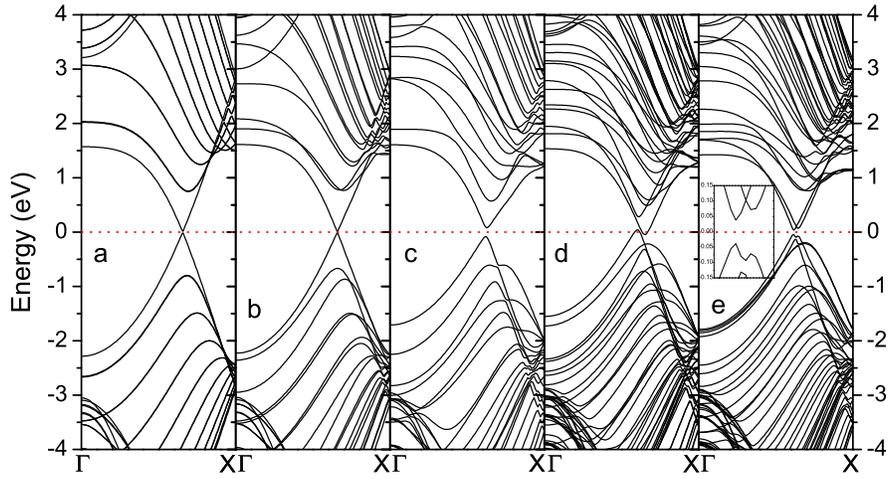}
\label{fig2} \caption{(Color online) The calculated electronic band
structures along the nanotube symmetry axis \textit{$\Gamma $X} of
(a) an individual (10,10) SWNT at 0 GPa, (b) a (10,10) SWNT bundle
at 0 GPa, and of three collapsed structures at 3 GPa: (c) the
parallel structure, (d) the in-between structure, (e) the
herringbone structure. Inset of (e) shows the pseudo-gap at the
Fermi level in an energy region of -0.15 $\sim$ 0.15 eV. The Fermi
level is set at zero.}
\end{figure}

\begin{figure}[htbp]
\includegraphics[width=0.8\columnwidth]{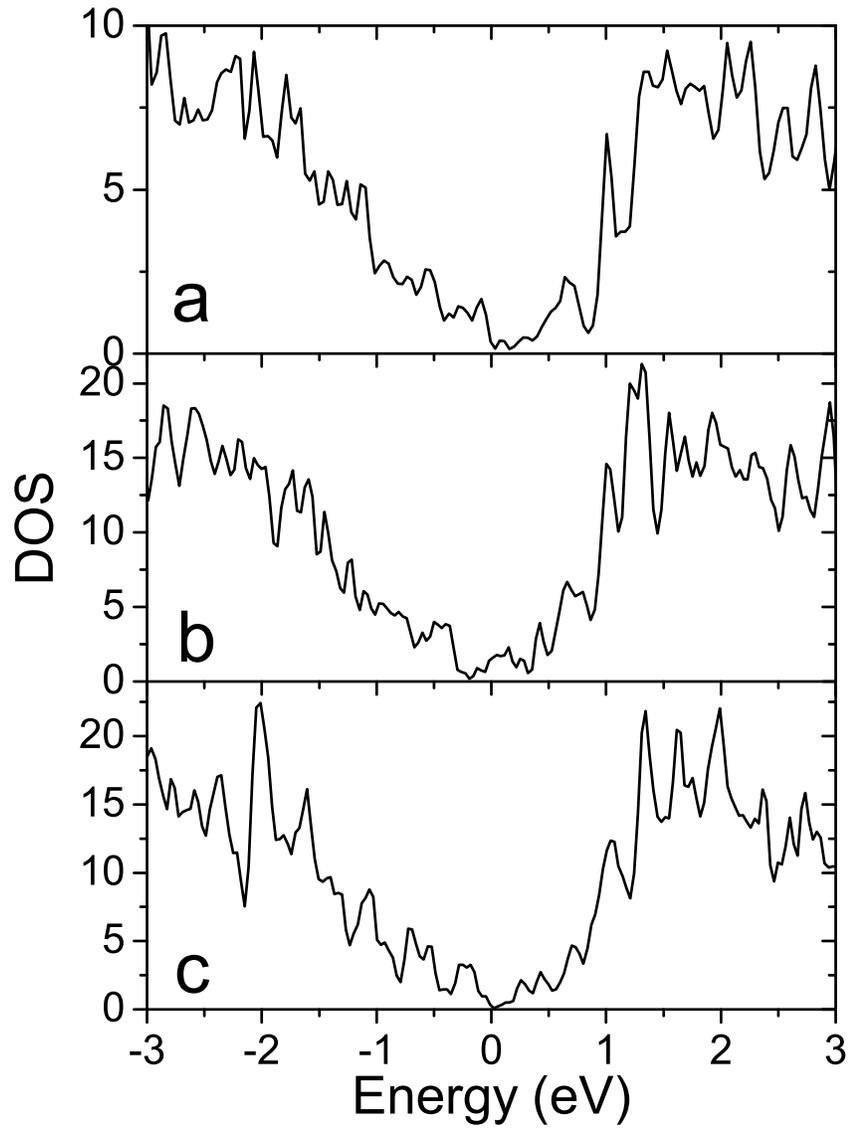}
\label{fig3} \caption{The calculated density of states for three
collapsed structures at 3 GPa: (a) the parallel structure, (b) the
in-between structure, and (c) the herringbone structure. The Fermi
level is set at zero.}
\end{figure}

\begin{figure}[htbp]
\includegraphics[width=0.8\columnwidth]{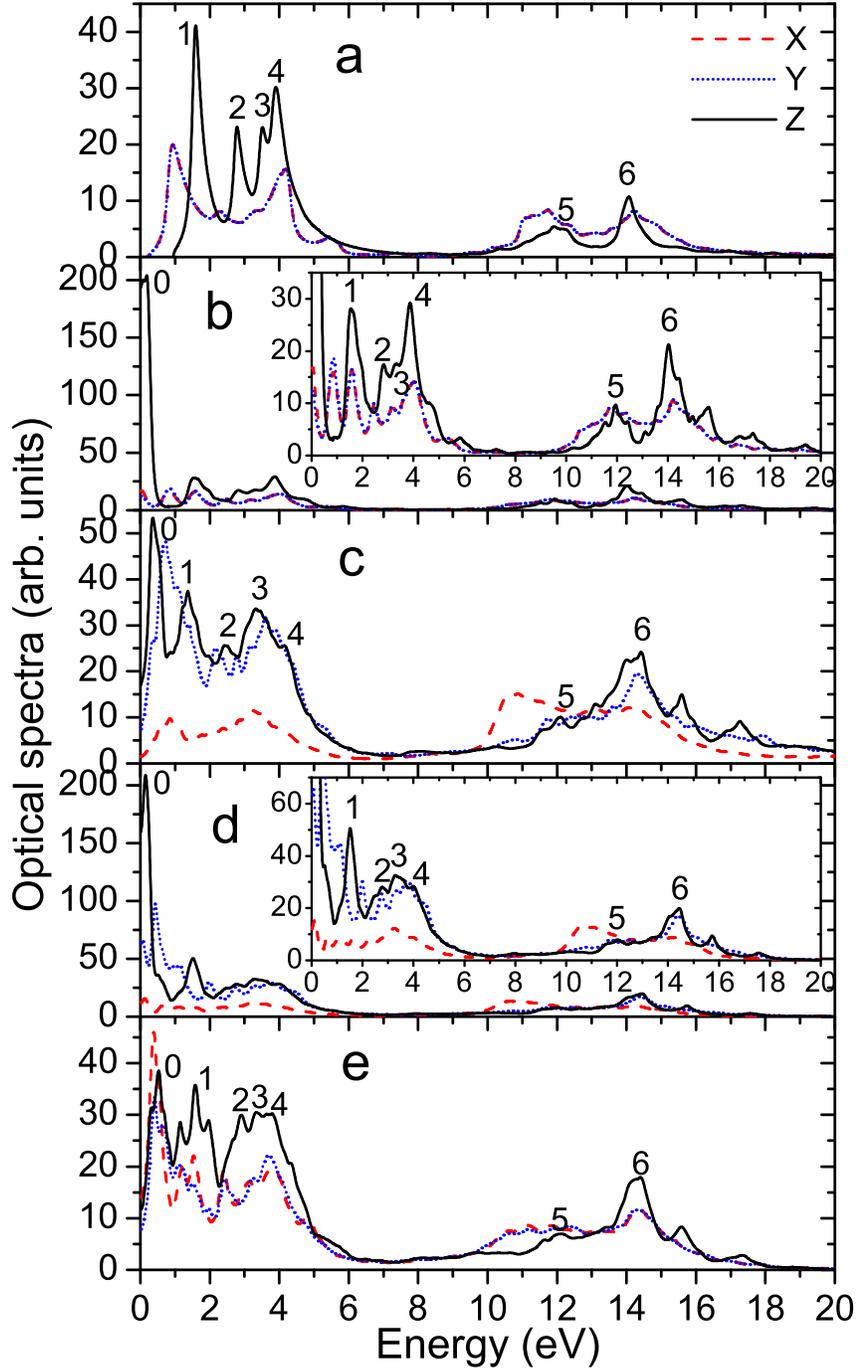}
\label{fig4} \caption{(Color online) The calculated imaginary part
(absorptive part) of dielectric function polarized along X, Y and Z
directions, respectively, for (a) an individual (10,10) SWNT at 0
GPa, (b) a (10,10) SWNT bundle at 0 GPa, and for three collapsed
structures at 3 GPa: (c) the parallel structure, (d) the in-between
structure, and (e) the herringbone structure.}
\end{figure}

\begin{figure}[htbp]
\includegraphics[width=0.7\columnwidth]{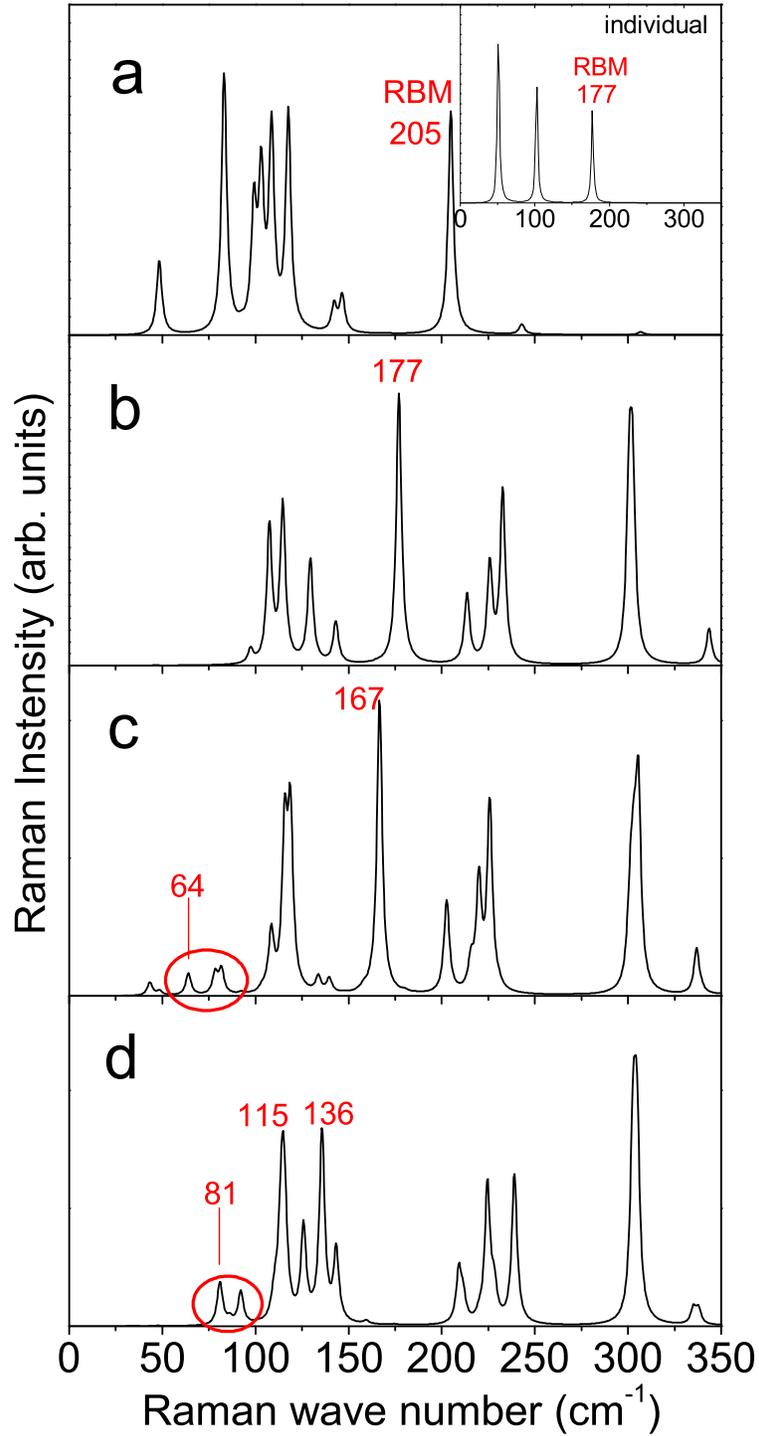}
\label{fig5} \caption{(Color online) The calculated nonresonant
Raman spectra of (a) a (10,10) SWNT bundle at 0 GPa (its inset shows
that of an individual (10,10) SWNT), and of three collapsed
structures at 3 GPa: (b) the parallel structure, (c) the in-between
structure, and (d) the herringbone structure.}
\end{figure}

\begin{figure}[htbp]
\includegraphics[width=0.8\columnwidth]{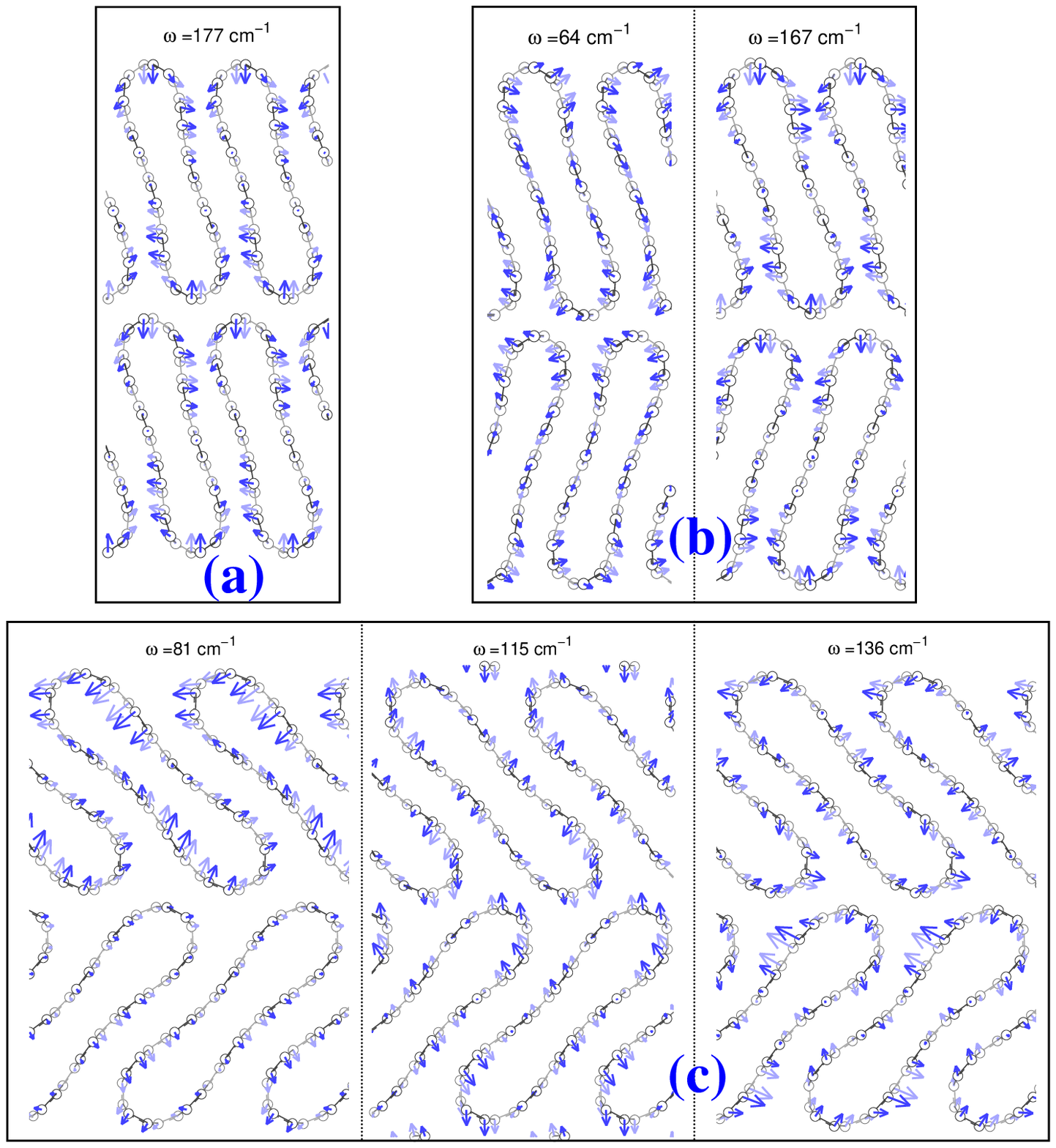}
\label{fig6} \caption{(Color online) The eigenfrequencies and
eigenvectors of some raman-active vibration modes for the different
collapsed structures at 3 Gpa: (a) the parallel structure, (b) the
in-between structure, and (c) the herringbone structure. The small
circles represent the carbon atoms and the straight lines between
them indicate the bonds. The arrows represent the atomic motions.
Deeper color arrow means the corresponding atom is nearer to the
reader.}
\end{figure}

\end{document}